\documentclass[prd,aps,twocolumn,a4paper,floatfix,nofootinbib]{revtex4}

\usepackage{graphicx,psfrag}
\usepackage{mathrsfs}
\usepackage{amsmath,amsfonts,amssymb}
\usepackage{multirow}
\usepackage{comment}
\usepackage{hyperref}
\usepackage{bbm}
\usepackage{placeins}
\usepackage[normalem]{ulem}
\usepackage{slashed}
\usepackage{amsthm}
\usepackage{latexsym}
\usepackage[dvips]{epsfig}
\usepackage{wasysym}
\usepackage{bm}
\usepackage{yhmath}
\usepackage{wrapfig}
\usepackage{stmaryrd}
\usepackage{pifont} 
\usepackage[utf8]{inputenc}

\def\ul{\underline}
\def\p{\partial}

\allowdisplaybreaks

\theoremstyle{plain}

\begin{document}

\title{The Hyperboloidal Numerical Evolution of a Good-Bad-Ugly Wave
  Equation}

\author{Edgar \surname{Gasper\'in}$^{1}$, Shalabh Gautam$^{2}$, David
  Hilditch$^{1}$ and Alex Va\~n\'o-Vi\~nuales$^{1,3}$}

\affiliation{ ${}^1$CENTRA, Departamento de F\'isica, Instituto
  Superior T\'ecnico IST, Universidade de Lisboa UL, Avenida Rovisco
  Pais 1, 1049 Lisboa, Portugal,\\ ${}^2$Inter-University Centre for
  Astronomy and Astrophysics, Post Bag 4, Ganeshkhind, Pune 411007,
  India,\\ ${}^3$School of Physics and Astronomy, Cardiff University,
  Queen’s Buildings, The Parade, CF24 3AA, Cardiff, United Kingdom }

\date{\today}

\begin{abstract}
One method for the numerical treatment of future null-infinity is to
decouple coordinates from the tensor basis and choose each in a
careful manner. This dual-frame approach is hampered by
logarithmically divergent terms that appear in a naive choice of
evolved variables. Here we consider a system of wave equations that
satisfy the weak-null condition and serve as a model system with
similar nonlinearities to those present in the Einstein field
equations in generalized harmonic gauge. We show that these equations
can be explicitly regularized by a nonlinear change of
variables. Working in spherical symmetry, a numerical implementation
of this model using compactified hyperboloidal slices is then
presented. Clean convergence is found for the regularized
system. Although more complicated, it is expected that general
relativity can be treated similarly.
\end{abstract}

\maketitle

\section{Introduction}\label{Section:Introduction}

For applications in gravitational wave astronomy it is desirable to
consider generic asymptotically flat spacetimes and compute, using the
methods of numerical relativity, waveforms at infinity. This is a
long-standing open problem. Following Penrose~\cite{Pen63} and
Friedrich~\cite{Fri81,Fri81a}, a natural strategy is to work on
compactified hyperboloidal slices, see~\cite{Zen08} or~\cite{Fra04}
for reviews. One proposal for this problem~\cite{HilHarBug16} is to
use a dual-frame (DF)~\cite{Hil15} approach. In this setting the
Einstein field equations (EFEs) are written using a generalized
harmonic tensor basis, and solved in the aforementioned hyperboloidal
coordinates. The key necessary requirement for this approach to work
is that certain derivatives of outgoing radial coordinate lightspeeds
have suitable decay. Recently it was shown~\cite{GasHil18} that this
{\it lightspeed condition} is related to the weak-null
condition~\cite{LinRod03}, an important structural condition for
small-data global existence to nonlinear wave-equations,
see~\cite{Kei17} for further discussion of the latter.  Using the
notion of the asymptotic system, it was furthermore heuristically
shown that even in a pure free-evolution setup, by making a suitable
addition of constraints to the field equations, the lightspeed
condition can be satisfied within the scheme of~\cite{HilHarBug16}.

The equations of motion in~\cite{HilHarBug16} were constructed to
avoid the presence of formally singular terms, but the simplest choice
of variables leaves some variables like~$(\ln R)/R$, near
null-infinity. Such terms could be problematic for
numerics. Therefore, going beyond the bare-bones scheme
of~\cite{HilHarBug16}, we would like to obtain {\it regular equations
  for regular unknowns that are not required to vanish at future-null
  infinity}. The aim here is to show that this can be achieved in a
nonlinear {\it good-bad-ugly} wave equation model whose
nonlinearities mimic those present in the formulation of the EFEs
given in \cite{GasHil18}. This builds towards a full-blown
regularization of general relativity (GR) in harmonic gauge, which
will be presented elsewhere, and may serve as an alternative to the
conformal field equations~\cite{Val16}. We present an implementation
with a battery of tests indicating that reliable, convergent, results
can be obtained using our regularization technique.

\section{A good-bad-ugly model}

In this section we present our model problem and give a new strategy
for regularization on hyperboloidal slices that exploits the nonlinear
structure of the field equations. We start in
section~\ref{Section:good-bad-ugly-secondorder} with the model in
second order form and then reduce to first order in
section~\ref{Section:semilineartoy}. The regularization is implemented
in section~\ref{Section:EvoEqsForTheRescaledVariables}, and finally
give a form of the asymptotic system in compactified hyperboloidal
coordinates in section~\ref{Section:CompactifiedAsymptoticSystem}.

\subsection{Second order form}\label{Section:good-bad-ugly-secondorder}

In our previous study~\cite{GasHil18} we found that in harmonic gauge
the field equations of GR can be divided into three categories, which
we call the good, the bad and the ugly. This categorization is made by
deriving from the original set of equations an associated asymptotic
system, which in turn can be used to predict decay rates of the
variables near null-infinity. The good equations are those for whom
the asymptotic system indicates fall-off identical to that of the
standard wave-equation. The bad equations are those whose solutions
decay slower than that. In the case of GR in harmonic gauge this can
be restricted to a logarithmic loss in a single equation. The ugly
equations are those associated with the constraints. These equations
can be manipulated by addition of the constraints. This can give
messier expressions, but ultimately results in fall-off faster than
that of the wave equation. Let~$(\mathbb{R}^4, m_{ab})$ denote the
Minkowski spacetime and metric and denote its corresponding
Levi-Civita connection as~${}{\nabla}$. The model equation to be
studied in this paper,
\begin{flalign}\label{eqn:ToyModelI}
\square g=0, \qquad {}{\square}b=(\p_T g)^2, \qquad
\square u=\tfrac{2}{\chi}\p_T u\,,
\end{flalign}
consists of a system of wave equations that ape this
structure. Here~${}{\square}$ denotes the d'Alembert operator in the
Minkowski spacetime and ~$\chi \simeq R$
where~$R^2=\delta_{\ul{ij}}X^{\ul{i}}X^{\ul{j}}$. In these
expressions~$X^{\ul{\mu}}=(T,X^{\ul{i}})$,
with~$\ul{\mu}\in\{0,1,2,3\}$ and~$\ul{i}\in\{1,2,3\}$, denote global
inertial Cartesian coordinates. We use the symbol~$\simeq$ to
represent equality at large radius up to error terms that decay faster
in~$R$ than those displayed in the expression. For instance~$f=R^{-1}
+O(R^{-(1+\delta)})$ with~$\delta>0$ will be written simply as
$f\simeq R^{-1}$. The model~\eqref{eqn:ToyModelI} is an example of a
system of wave equations satisfying the weak null condition. To
understand what this means, one has to derive the asymptotic system
mentioned above.

Here we sketch the construction as given in~\cite{LinRod03}. In
section~\ref{Section:CompactifiedAsymptoticSystem} we give an
alternative method to obtain the asymptotic system that uses
hyperboloidal coordinates directly. We begin by introducing
shell-coordinates~$X^{\underline{\mu'}}=(T,R,\theta^A)$
where~$\theta^A$ with~$A \in \{1,2\}$ denote arbitrary coordinates
on~$\mathbb{S}^2$, whose specific form will be fixed later, and then
defining the rescaled variables~$\mathcal{G}= R g$, $\mathcal{B} = R b
$ and~$ \mathcal{U}= R u$. Now, consider the coordinate
change~$(T,R,\theta^A) \rightarrow (\upsilon,s,\theta^A)$ where
$\upsilon=T-R$ and $s= \ln R$. Rewriting the system in terms
of~$\mathcal{G}$, $\mathcal{B}$ and~$\mathcal{U}$ and formally
equating the terms with coefficients $1/R^2$ gives,
\begin{align}
 2\p_s\p_\upsilon \mathcal{B} = -(\p_\upsilon\mathcal{G})^2, \quad
\p_s\p_\upsilon \mathcal{G}=0, \quad \p_s\p_\upsilon
\mathcal{U}=-\p_\upsilon \mathcal{U}.\label{AsymptoticToySecondOrder}
\end{align}
This is the asymptotic system associated with
equation~\eqref{eqn:ToyModelI}. Observe that the second equation
in~\eqref{AsymptoticToySecondOrder} implies that~$\p_\upsilon
\mathcal{G}$ depends only on~$\upsilon$. Consequently, one can
integrate the first equation of~\eqref{AsymptoticToySecondOrder}
in~$s$ and conclude that~$\p_\upsilon \mathcal{B} = -\tfrac{s}{2}
(\p_\upsilon{} \mathcal{G})^2$. Hence in the asymptotic
approximation~$\mathcal{B} =
-\tfrac{1}{2}\ln(R)\int_{\upsilon_\star}^\upsilon
(\p_{\bar{\upsilon}}\mathcal{G})^2\mbox{d}\bar{\upsilon}$. From
the third equation one concludes that~$\p_\upsilon \mathcal{U} =
\frac{1}{R}(\p_\upsilon \mathcal{U})|_{s_\star}$. Integrating
this gives~$\mathcal{U}=\frac{1}{R}\int_{\upsilon_\star}^\upsilon
(\p_\upsilon \mathcal{U})|_{s_\star}\mbox{d}\bar{\upsilon}$. So
the asymptotic system for~$u$ indicates that one could have taken a
further power of~$1/R$ in the definition of the rescaled
variable~$\mathcal{U}$. [The right-hand side of the equation for~$u$
  was of course chosen precisely for this purpose]. Nevertheless,
observe that given a function~$m=m(\theta,\varphi)$ such that
$\slashed{\triangle}m=0$, where~$\slashed{\triangle}$ denotes the
Laplacian on~$\mathbb{S}^2$, one has that $m/R$ is a solution to any
of the equations in \eqref{eqn:ToyModelI}. In the remainder we will
discard these solutions. In other words, we consider fields $u$, $g$
and $b$ where we have subtracted any \emph{static} solutions of this
form.

A system of wave equations is said to satisfy the weak null condition
if its asymptotic system admits global solutions which grow at most
exponentially in~$s$~\cite{LinRod03}. Recall that a quadratic
form~$N^{\mu\nu}\p_\mu\phi\p_\nu \phi$ is a null-form if it vanishes
upon replacement of $\p_\mu\phi$ with $\xi_\mu$ where $\xi^\mu$ is a
null vector. A wave equation is said to satisfy the classical null
condition~\cite{Kla80, Kla86, Chr86} if its nonlinearity can be
written as a sum of null-forms. A system of quasilinear wave
equations satisfying the classical null condition trivially satisfies
the weak null condition. One naive question that arises from this
analysis is whether the logarithmic loss predicted by the asymptotic
equation for~$\mathcal{B}$ is somehow tied to the choice of the
coordinate system~$(\upsilon,s,\theta^A)$. In other words, is the
logarithmic loss only present due to the choice~$s(R)=\ln(R)$, or
could better coordinates give a sharper result? In
Section~\ref{Section:CompactifiedAsymptoticSystem} a new version of
the above analysis is given. In this new approach a compactified
radial coordinate is used instead of the aforementioned {\it
  stretch}~$R=e^s$. Identical results are obtained, indicating that
the outcome is robust against coordinate changes.

The model equation~\eqref{eqn:ToyModelI} is an example of a system of
wave equations that fails to satisfy the classical null condition, but
does satisfy the weak null condition.  We call the system of
equations~\eqref{eqn:ToyModelI} the semilinear good-bad-ugly model
since~$g$ is a field (the good) that satisfies an equation that
fulfills the classical null condition while~$b$ (the bad) is a field
which is responsible for the failure of the system to satisfy the
classical null condition. Since the Einstein field equations in
harmonic gauge satisfy the weak null condition the good-bad subsystem
alone constitutes a simple toy model to be implemented numerically in
the dual foliation framework~\cite{HilHarBug16, Hil15} which
en-captures this feature. This subsystem is furthermore the simplest
toy model in which one can test regularization strategies for the
fields analogous to the bad metric components, as they appear in
generalized harmonic gauge in the EFE. The addition of the
variable~$u$ (the ugly) to the good-bad subsystem arises as a model
for the type of equations that appear in the evolution equations if
one adds certain multiples of the harmonic constraints to the EFE. As
discussed in~\cite{GasHil18} such an addition is required for a
successful numerical implementation of the hyperboloidal initial value
problem using the DF formalism.  The equations in
expression~\eqref{eqn:ToyModelI} almost decouple, naturally we could
have analyzed a more complicated system of equations satisfying the
weak null condition. Adding null-forms for instance would leave the
asymptotic system unchanged, but we instead want to consider the
simplest good-bad-ugly model with the properties described above.

\subsection{First order reduction of the model}\label{Section:semilineartoy}

Let~$(\mathbb{S}^2, q_{ab})$ denote the unit two-sphere with the
standard metric and represent by~${}{\slashed{\nabla}}$ the associated
Levi-Civita connection. The shell
coordinates~$X^{\ul{\mu'}}=(T,R,\theta^{A})$ will be regarded, in the
language of the DF formalism~\cite{Hil15} as the {\it uppercase}
coordinates.

For numerical implementation it is convenient to perform a first order
reduction of the model. To that end, we make use of the usual~$3+1$
decomposition. Let~$\Sigma_T$ be the hypersurface determined by the
condition~$T=\textrm{const.}$,~${}{N}^a$ denotes the timelike unit
normal to this surface and define the projector~$\gamma_{ab}$
as~$^{(N)}\gamma_{ab}=m_{ab}+{}{N}_{a}{}{N}_{b}$. We then define the
time reduction variable~$g_\pi$ by~$g_\pi=-\p_Tg$ and introduce
a spatial covector~$g_a$ and denote as~$g_R$ and~$g_{A}$ the
components of~$g_a$ respect
to~$((\p_R)^a,(\p_{A})^a)$. The reduction constraints are
\begin{align}\label{ReductionConstraintToyModel}
  {}^{(g)}\mathcal{C}_a \equiv \gamma_a{}^{b}{}{\nabla}_b g - g_a.
\end{align}
Analogous definitions are made for the~$b$ and~$u$ fields. Using this
notation, the semilinear good-bad-ugly model system can be written in
first order form as follows. For the goods we have,
\begin{subequations}
  \label{eq:firstorderevolutionequationslowercasecoords}
   \begin{align}
  \p_T g  &=  - g_{\pi}, \label{tildephi}\\
  \p_T g_{R}  &=  -   \p_Rg_\pi +
  \gamma_{2} (- g_{R}{}
   + \p_Rg), \label{tildephiR}\\
   \p_T g_{A}  &=  - \p_{A}g_\pi +
  \; \gamma_{2} (- g_{A}{} +  \p_{A}g), \label{tildephiA}\\
  \p_T g_{\pi}  &=   -\p_R g_{R}
  -\tfrac{1}{R^2}q^{AB}\slashed{\nabla}_{A}g_{B}
  -\tfrac{2}{R}g_{R}, \label{tildepi}
   \end{align}
while for the bads the equations are
   \begin{align}
  \p_T b  &=  - b_\pi, \label{notildephi}\\
  \p_T b_{R}   &=  -  \p_R\pi + \gamma_{2}{} (-b_{R}{} +
  \p_R b), \label{notildephiR}\\
 \p_T b_{A}  = &  - \p_{A}b_\pi +
  \; \gamma_{2} (- b_{A}{} +  \p_{A}b), \label{notildephiA}\\
  \p_T b_{\pi}  &=  \; g_\pi{}^2 -\p_R b_{R}
  -\tfrac{1}{R^2}q^{AB}\slashed{\nabla}_{A}b_{B}
  -\tfrac{2}{R}b_{R},\label{notildepi}
   \end{align}
and finally for the uglies, the evolution equations read
  \begin{align}
   \p_T u  &=  - u_{\pi}, \label{ugly}\\
  \p_T u_{R}  &=  -   \p_Ru_\pi +
  \gamma_{2} (- u_{R}{}
   + \p_Ru), \label{uglyR}\\
   \p_T u_{A}  &=  - \p_{A}u_\pi +
   \; \gamma_{2} (- u_{A}{} +  \p_{A}u), \label{uglyA}\\
   \p_T u_{\pi}  &= -\tfrac{2}{\chi}u_{\pi}
   -\p_R u_{R} -\tfrac{1}{R^2}q^{AB}\slashed{\nabla}_{A}u_{B}
   -\tfrac{2}{R}u_{R} . \label{uglypi}
\end{align}
\end{subequations}
Here~$\gamma_2$ is a freely prescribable scalar function of the
coordinates. The definition of the time reduction variables is encoded
in the evolution equations~\eqref{tildephi}, \eqref{notildephi}
and~\eqref{ugly}. Setting~$\gamma_2=0$ one sees that the evolution
equations~\eqref{tildephiR}, \eqref{notildephiR} and~\eqref{uglyR}
were obtained from the no-torsion
condition~$[{}{\nabla}_a,{}{\nabla}_b]\phi = 0$, with~$\phi=g,b,u$.
Observe that the term next to~$\gamma_2$ corresponds to the reduction
constraints. These terms are introduced to mitigate constraint
violation in free evolution schemes. Their addition affects the
principal part of the equations, but nevertheless the system is
symmetric hyperbolic for any choice of the formulation
parameter~$\gamma_2$~\cite{LinSchKid05}. Finally,
expressing~\eqref{eqn:ToyModelI} using the reduction variables renders
the evolution equations~\eqref{tildepi}, \eqref{notildepi}
and~\eqref{uglypi}.

\subsection{Evolution equations for the rescaled variables}
\label{Section:EvoEqsForTheRescaledVariables}

In this section we rescale the variables and derive evolution
equations for them. In the construction of the asymptotic system we
rescaled all of the fields identically. As examined
in~\cite{GasHil18}, this leads to a first order version of the
asymptotic system discussed in
section~\ref{Section:good-bad-ugly-secondorder}. Nevertheless,
evolving the first order variables analogous to~$\mathcal{G}$,
$\mathcal{B}$ and~$\mathcal{U}$ is not convenient for numerical
applications because~$\mathcal{B}$ is expected to diverge as~$\ln(R)$
and~$\mathcal{U}$ to decay as~$1/R$ near null-infinity. Ideally we
want regular equations for variables for which one expects a regular
behavior and a finite, potentially non-zero, limit at null
infinity. The latter suggests performing a different rescaling for the
good, the bad and the ugly fields. In this section we discuss how to do
this in such a way that the evolution equations for the rescaled
variables in hyperboloidal coordinates~$x^\mu=(t,r, \theta^A)$,
corresponding to the \emph{lowercase coordinates} in the nomenclature
of the DF formalism, are regular.

\paragraph*{Asymptotic system primer:} We denote the outgoing and
incoming null vectors by~$L$ and~$\underline{L}$ respectively,
\begin{align}\label{eq:NullFrameToSpFrame}
L^a &=  \; (\p_T)^a + (\p_R)^a, \qquad
\underline{L}^a=(\p_T)^a - (\p_R)^a,
\end{align}
and define
\begin{align}\label{characteristicvariablesg}
g^+&= -g_\pi+g_R, \quad g^-=-g_\pi-g_R,
\end{align}
and analogous quantities for the fields~$b$ and~$u$. Observe that
the~`$+$' and~`$-$' variables correspond to the~$L$
and~$\underline{L}$ derivatives of the original fields, or, in other
words, the characteristic variables of the
system~\eqref{eq:firstorderevolutionequationslowercasecoords}
with~$\gamma_2=0$. Substituting~$g_\pi$, $g_R$, $b_\pi$, $b_R$,
$u_\pi$ and~$u_R$ written in terms of the~`$+$' and~`$-$' variables
into equations~\eqref{tildephi}-\eqref{uglypi} one obtains a set of
evolution equations for the~`$+$', `$-$' and~`$_A$' variables for~$g$,
$b$ and~$u$. The next step is to rescale and choose the variables to
evolve. Since there is a large freedom in doing this in practice, to
explain the rationale behind the choice that we make, first, let us
examine the standard rescaling suggested by the discussion of the
asymptotic system of Section~\ref{Section:good-bad-ugly-secondorder}. We
use a schematic notation and let~$\phi$
represent~$g,b,u$. Likewise~$\phi_A$ encodes the angular variables
and, as in equation~\eqref{characteristicvariablesg},~$\phi^{\pm}$ the
characteristic variables.  Similarly we denote by~$\Phi, \Phi_A,
\Phi^{\pm}$ the rescaled variables~$\mathcal{G}, \;\mathcal{G}_A,\;
\mathcal{G}^{\pm},\; \mathcal{B},\; \mathcal{B}_A,\;
\mathcal{B}^{\pm},\; \mathcal{U},\; \mathcal{U}_A,\;
\mathcal{U}^{\pm}$.  With this notation at hand, the following
expressions should be regarded as the `basic' rescaling,
\begin{equation}\label{BasicRescaling}
\Phi^{+}= R^2\phi^+, \;\;\Phi^-=R\phi^-, \;\; \Phi = R \phi,\;\;
\Phi_A = R\phi_A\,.
\end{equation}
This corresponds, in terms of the original fields to taking the
derivative first and rescaling afterwards, namely
\begin{equation}\label{BasicRescalingAlt}
\Phi^{+}= R^2L\phi, \;\;\Phi^-=R\underline{L}\phi, \;\; \Phi = R \phi,\;\;
\Phi_A = R\p_A\phi.
\end{equation}
Following the procedure described in~\cite{GasHil18} and, for
simplicity of the presentation, taking~$\gamma_2=0$, one obtains the
following first order asymptotic system:
\begin{align}\label{AsymptoticSystemGBUstandardRescaling}
  \p_\upsilon \mathcal{G}^+ &= -\tfrac{1}{2}\mathcal{G}^{-}\,, \qquad &
  \p_\upsilon \mathcal{B}^+ &=
  -\tfrac{1}{2}\mathcal{B}^{-}-\tfrac{1}{8}(\mathcal{G}^-)^2\,, \nonumber
  \\ \p_s \mathcal{G}^{-} &= 0\,, \qquad & \p_s \mathcal{B}^{-} &=
  -\tfrac{1}{4}(\mathcal{G}^-)^2\,, \nonumber \\
  \p_\upsilon\mathcal{U}^+ &= -\mathcal{U}^{-}\,, \qquad & \p_\upsilon \Phi & =
  \tfrac{1}{2}\Phi^{-}\,, \nonumber \\ \p_s \mathcal{U}^{-} &=
  -\mathcal{U}^{-} \qquad & \p_\upsilon \Phi_A & = \tfrac{1}{2}\p_A
  \Phi^{-}\,.
\end{align}
Since we have chosen here the same rescaling for all the variables
regardless of the equation they satisfy it is natural that their
asymptotic equations differ. Nonetheless, in the following we will
discuss how to exploit the information provided by the asymptotic
system~\eqref{AsymptoticSystemGBUstandardRescaling} in order to obtain
optimal definitions for the rescaled variables.

First notice that, although the decay of the good fields cannot be
improved, the asymptotic system for these fields can be written in a
slightly simpler way by considering the following variables
\begin{equation}\label{Grescling}
G^{+}= \mathcal{G}^+ +\mathcal{G}, \;\; G^-=\mathcal{G}^-, \;\; G =
\mathcal{G},\;\; G_A = \mathcal{G}_A.
\end{equation}
Then, the asymptotic system for the good variables reads,
\begin{align}\label{prototypeasymptoticsys}
  \p_\upsilon G^+ &= 0, & \qquad \p_s G^{-} &= 0, \nonumber
  \\ \p_\upsilon G & = \tfrac{1}{2}G^{-}, & \p_\upsilon G_A & =
  \tfrac{1}{2}\p_A G^{-}.
\end{align}
In terms of the original unrescaled field, the latter change
corresponds to the following choice of variables.
\begin{equation}\label{DemystifiedGrescaling}
G^{+}= RL(Rg), \;\;G^-=R\underline{L}g, \;\; G = R g,\;\;
G_A = R\p_Ag.
\end{equation}
Henceforth, we regard the form of the
equations~\eqref{prototypeasymptoticsys} as the prototype for the
asymptotic system for a set of equations that have the same
asymptotics as that of the homogeneous wave equation.

In order to find the optimal redefinition for the rescaled uglies,
first recall that, as already discussed in
Section~\ref{Section:good-bad-ugly-secondorder}, the in-homogeneity in
the equation for the ugly was designed so that the field~$u$ decays
one order~$1/R$ faster than~$g$. This property can be read off
directly from the asymptotic equation~$\p_s \mathcal{U}^{-} =
-\mathcal{U}^{-}$. This suggests rescaling the fields with one extra
power of~$1/R$ of that of the basic rescaling~\eqref{BasicRescaling},
furthermore, some experimentation reveals that by defining
\begin{align}
U^{+}&= R(\mathcal{U}^{+} + 2\mathcal{U}), &\quad U^-&=R\mathcal{U}^-,\nonumber\\
U&= R \mathcal{U},&\quad U_A &= R \mathcal{U}_A,\label{Urescling}
\end{align}
the asymptotic equations for these variables are identical
to~\eqref{prototypeasymptoticsys} if we-formally replace~$G^{\pm}$
and~$G$ by~$U^{\pm}$ and~$U$ respectively. Written in terms of the raw
ugly field we have,
\begin{align}
U^{+}&= RL(R^2u), &\quad U^-&=R^2\underline{L}u, \nonumber\\
U &= R^2 u,&\quad
U_A&= R^2\p_Au.\label{DemystifiedUrescaling}
\end{align}
For the bad fields we make the following nonlinear change of
variables
\begin{align}\label{Brescling}
    B^+ &= \mathcal{B}^+ + \mathcal{B} + \tfrac{1}{8}\eta, & \quad B^-
    & = \mathcal{B}^- +\tfrac{1}{4}s\p_\upsilon \eta, & \nonumber \\ B
    & = \mathcal{B} + \tfrac{1}{8}s \eta, & \quad B_A & = \mathcal{B}_A
    +\tfrac{1}{8}\p_A \eta. &&
\end{align}
where~$\eta$ is an auxiliary variable whose evolution equation has to
be chosen in such a way that in the asymptotic limit it reduces
to~$\p_\upsilon \eta = (G^-)^2$. With these definitions, a direct
calculation shows that the asymptotic system for the~$B$-fields, which
we call the \emph{reformed bads} is identical to that of the goods
under replacement of~$G^{\pm} $ and~$G$ by~$B^{\pm}$ and~$B$
respectively in equation~\eqref{prototypeasymptoticsys}. The price to
pay for this regularization is the introduction of a new
variable~$\eta$ which, as in the asymptotic system,
encodes~$\int_{\upsilon_\star}^\upsilon(\p_{\bar{\upsilon}}G)^2
\mbox{d}\bar{\upsilon}$. Although~$\eta$ is defined as an integral it
will satisfy a local equation of motion. The change of
variables~\eqref{Brescling} can be written in terms of the raw bad
field as
\begin{align}
  B^{+} & = RL(Rb + \tfrac{1}{8}s\eta), & B^- & = R\underline{L}(b)
  + \tfrac{1}{8}s\underline{L}\eta, \nonumber\\
  B & = R b + \tfrac{1}{8}s \eta, &
B_A & = R\p_Ab + \tfrac{1}{8}s\p_A\eta.\label{DemystifingBrescling}
\end{align}
This regularization strategy can be thought of as ``subtracting''
the~$\ln (R)$ part of the asymptotic solution for~$\mathcal{B}$. An
alternative regularization strategy is, instead, to ``divide'' by~$\ln
(R)$. The disadvantage of the latter, arguably simpler option, is that
it generates slowly decaying~$1/\ln(R)$ terms in the evolution
equations. We have implemented this regularization also, but find that
these slowly decaying terms prevent the code from converging, and so
do not discuss the method further.

\paragraph*{Complete evolution equations in hyperboloidal coordinates:}
The foregoing discussion already demonstrates how to choose the
rescaled variables in order to have regular equations
at~$\mathscr{I}^+$. Nevertheless, as will be elaborated further in the
remainder of this section, care is needed at the origin~$R=0$ if we
wish to evolve numerically the good-bad-ugly system in spherical
symmetry. Taking this into account, a suitable choice for the rescaled
variables is, for the goods,
\begin{align*}
  G^+ =\; & \chi{}^2 g^+ + R g, &  G^- = \; & \chi{}g^-,
  \\ G = \; & \chi{} g, & G_{A} =\; & \chi {}g_{A},  \nonumber
\end{align*}
for the bads,
\begin{align*}
 B^+ = \; & \chi{}^2 b^+ + R b + \tfrac{1}{8}\chi^{-1}R\eta, & \;\;
 B = \; & \chi{} b  +  \tfrac{1}{8}(\xi-1)\eta, &  \nonumber \\
 B^- = \; & \chi{} b^- +  \tfrac{1}{4}(\xi-1)\p_T\eta, &  \;\;
 B_A = \; & \chi{} b_A  +  \tfrac{1}{8}(\xi-1)\eta_A,
\end{align*}
and finally, for the uglies,
\begin{align}
  U^{+} = \; & \chi{}^3 u^+ + 2 \chi R u, & U^- = \;
  & \chi^2 u^-,  \nonumber \\  U = \; & \chi{}^2 u,  & U_{A} =\; & \chi^2 u_{A},
  \label{eq:rescaling}
\end{align}
where~$\chi=\chi(R)$ and~$\xi \equiv \ln\chi$. The even
function~$\chi$ is to be chosen such that~$\chi(0)=1$ and~$\chi \simeq
R$ at large radii. This ensures that, asymptotically, the change of
variables is that of equations~\eqref{Grescling}, \eqref{Urescling}
and~\eqref{Brescling}, while at the origin the transformation reduces
to the identity. Additionally, we have introduced new variables~$\eta$
and~$\eta_A$, the latter encoding the angular derivatives of the
former. Thus, associated to~$\eta_A$ we introduce the constraint
\begin{align}
{}^{(\eta)}C  \equiv \slashed{\nabla}_{A} \eta - \eta_A\,.
\end{align}
Once the evolution equation for~$\eta$ is chosen, the equation
for~$\eta_A$ can be obtained exploiting the no-torsion
condition~$[\nabla_a, \nabla_b]\eta = 0$. From the previous discussion
we know that the evolution equation for~$\eta$ has to be chosen such
that it asymptotically reduces to~$\p_\upsilon \eta = (G^-)^2$ in
order for our regularization strategy to work. A simple choice, to
which we adhere from this point onward, is to set
\begin{align}
\p_{T}\eta = 4 R^2 g_{\pi}{}^2.
\end{align}
Expressed in rescaled variables this reads,
\begin{align}\label{evoEta}
  \p_{T}\eta = \tfrac{R^2}{\chi^2} \Bigl( G^{-}{} +
  \tfrac{1}{\chi} G^{+} - \tfrac{R}{\chi^2}G \Bigr)^2.
\end{align}
Using the no-torsion condition as described before and
equation~\eqref{evoEta} we obtain the following for~$\eta_A$,
\begin{align}
  \p_{T}\eta_{A}{} &= \tfrac{2 R^2}{\chi^2}
  \Bigl( G^{-}{} + \tfrac{1}{\chi} G^{+}{}
  - \tfrac{R}{\chi^2}G \Bigr)\times \nonumber\\
  &\quad\bigl( \slashed{\nabla}_{A}G^{-}{}
  + \tfrac{1}{\chi}\slashed{\nabla}_{A}G^{+}{}
  - \tfrac{R}{\chi^2} G_{A} \bigr)\,.
\end{align}
Expressing the evolution
equations~\eqref{eq:firstorderevolutionequationslowercasecoords} in
terms of the rescaled variables as defined around
equation~\eqref{eq:rescaling} is a straightforward but cumbersome
calculation. The reason for the latter is twofold: the change of
variables~\eqref{eq:rescaling} was designed so that at the origin the
rescaled variables reduce to the unrescaled characteristic
variables~$\phi^+,\phi^-, \phi$ and $\phi_A$. To do so, we had to
introduce functions such as~$\chi$ instead of simply~$R$ or~$\xi-1$
instead of just~$\ln(R)$, that when pushed through the change of
variables produce several non-principal terms. Second, although we
expect to obtain the simplest possible expression for the asymptotic
equation under this choice of variables, due to the extra term added
in the definition of the $+$ fields, we do not get simple advection
equations. Compare for instance the definition for~$G^+$
and~$\mathcal{G}^+$ in equations~\eqref{BasicRescaling}
and~\eqref{Grescling}. Nevertheless, we improve this situation by
adding multiples of the constraints appropriately to absorb these
extra terms and thus reduce the system to a set of advection equations
near infinity. We omit the details of this computation. We know that
the original equations are symmetric hyperbolic, but what of the
modified set? After changing variables, we end up with a system which
takes a standard first order symmetric hyperbolic form for all of the
fields except the reformed bads~$B,B^+,B^-,B_A$, plus~$\eta$
and~$\eta_A$, each of which look like a system that would be trivially
symmetric hyperbolic if the derivative coupling to the rescaled good
fields could be dropped. This additional coupling can be treated as
non-principal however, by noting that the full system can be
consistently evolved under the assumption that the good fields are one
degree of regularity better behaved (in suitable Sobolev spaces) than
the reformed bads. Although we have identified the leading behavior of
the fields via the asymptotic system analysis, a deeper understanding
of the solution could perhaps be achieved by obtaining a hierarchical
set of ``higher order asymptotic systems'' determining the subleading
terms in the solution.  This is left for future work. To express the
evolution equations in their final form, we define radially
compactified hyperboloidal coordinates~$(t,r,\theta^A)$ related
to~$(T,R,\theta^A)$ via,
\begin{align}
T &= t + H(R(r)),\qquad R=R(r),
\end{align}
and let~$H' = dH/dR$ and~$R'=dR/dr$. A direct computation shows that
the above evolution equations for the rescaled variables in
the~$(t,r,\theta^A)$ coordinates read as,
\begin{align}
\p_{t} G^{+}  &= -c_{-}^r\p_rG^{+}{}
- \mathcal{A}^-q^{AB}{\slashed{\nabla}}_A G_B + S_{G^+}, \nonumber \\
\p_{t} G^{-}  &= -c_+^{r}\p_rG^{-}{} -\gamma_2c_+^{r}\p_rG^{}
+\mathcal{A^{+}}q^{AB}{}{\slashed{\nabla}}_A G_B  + S_{G^-}, \nonumber\\
\p_tG_{A}{}  &= \tfrac{1}{2} \p_A (G^{-} + \tfrac{1}{\chi} G^{+} )
+ \gamma_2 \p_A G -\gamma_2G_A \nonumber \\
&\quad -\tfrac{R}{2\chi^2}G_A, \nonumber \\
\p_{t} G  &=  \tfrac{1}{2} G^{-} - \tfrac{R}{2\chi^2}G
+ \tfrac{1}{2\chi}G^{+},\label{eq:Recaled_Good_EoMs}
\end{align}
for the goods, whilst for the bads we have,
\begin{align}
\p_{t} B^{+}  &= -c_{-}^r\p_rB^{+} - \mathcal{A}^{-}q^{AB}{\slashed{\nabla}}_A B_B
+ F^{-}_{\eta}\p_r \eta \nonumber \\
&\quad + \tfrac{1}{8}(\xi-1)\mathcal{A}^{-}q^{AB}{\slashed{\nabla}}_A \eta_B
+ S_{B^+}, \nonumber \\
\p_{t} B^{-}   &= -c_+^{r}\p_rB^{-} -\gamma_2c_+^{r}\p_rB^{}
+\mathcal{A^{+}}q^{AB}{}{\slashed{\nabla}}_A B_B+F_{\eta}^{+}\p_r \eta \nonumber\\
&\quad - \tfrac{1}{8}(\xi-1)\mathcal{A}^+ q^{AB}{\slashed{\nabla}}_A \eta_B
+ M \p_r G^+ + \gamma_2J \p_r G \nonumber \\
&\quad + K q^{AB}{}{\slashed{\nabla}}_A G_B  + S_{B^-}, \nonumber \\
\p_tB_A & = \tfrac{1}{2} \p_A (B^{-} + \tfrac{1}{\chi} B^{+} )
+ \tfrac{1}{8}\gamma_2 (\xi-1) \p_A \eta + \gamma_2 \p_A B \nonumber \\
&\quad-\big( \gamma_{2} +  \tfrac{R}{2\chi^{2}}\big) B_A + F_\eta^\theta\eta_{A}{},
\nonumber\\
\p_{t} B  &=  \tfrac{1}{2} B^{-} - \tfrac{R}{2 \chi^{2}}B +
\tfrac{1}{2\chi}B^{+} + \tfrac{R}{16\chi^2}(\xi-2)\eta.\label{eq:Recaled_Bad_EoMs}
\end{align}
Note here the aforementioned derivative coupling to the rescaled good
fields, and the advection-equation like form of
both~\eqref{eq:Recaled_Good_EoMs} and~\eqref{eq:Recaled_Bad_EoMs}
near~$\mathscr{I}^+$.  Next for the uglies we get,
\begin{align}
\p_{t} U^{+}  &=  -c_{-}^r\p_rU^{+} - \mathcal{A}^-q^{AB}{\slashed{\nabla}}_A U_B +
S_{U^+}, \nonumber \\
\p_{t} U^{-}  &=  -c_+^{r}\p_rU^{-}{}-\gamma_2c_+^{r}\p_rU{}
+\mathcal{A^{+}}q^{AB}{}{\slashed{\nabla}}_A U_B  + S_{U^-}, \nonumber \\
\p_tU_{A}{}  &= \tfrac{1}{2} \p_A (U^{-} + \tfrac{1}{\chi} U^{+} )
+ \gamma_2 \p_A U -\gamma_2U_A -\tfrac{R}{\chi^{2}}U_A, \nonumber\\
\p_{t} U &=  \tfrac{1}{2} U^{-} - \tfrac{R}{2 \chi^{2}}U
+\tfrac{1}{2\chi}U^{+},
\end{align}
and for the auxiliary variable,
\begin{align}
\p_t\eta  &= \tfrac{R^2}{\chi^2}P^2 &\quad
\p_t\eta_A &= \tfrac{2 R^2}{\chi^{2}} P\p_A P,\label{eqn:evofinal}
\end{align}
where the various coefficients
in~\eqref{eq:Recaled_Good_EoMs}-\eqref{eqn:evofinal} are given by,
\begin{align*}
\alpha_{+} &= \; 1, & \alpha_{-} &= \; \chi, \\ \beta_{+} &= \; 0, &
\beta_{-} &= \; 1, \\ c^r_{\pm} &= \tfrac{\pm1}{(1 \mp H') R'}, &
\mathcal{A}^{\pm} &= \tfrac{R'}{R^2} \alpha_{\pm}c^r_{\pm},\\ M &=
\tfrac{ R' R^2}{\chi^3} (\xi-1) c_{-}^{r} c_{+}^{r}P, & K &=
\tfrac{R'}{R^{2}}\chi M, \\ J &= \tfrac{\chi(1-R'c_{+}^r)}{R'c_{+}^r}
M, & P &= G^{-} + \tfrac{1}{\chi} G^+ - \tfrac{R}{\chi^{2}}G,
\end{align*}
and,
\begin{align*}
F_\eta^\pm &= \tfrac{c_{\pm}}{8}\Big(\tfrac{R}{\chi}\beta_{\pm}
+\gamma_2(\xi-1)\alpha_{\pm}\Big),\\
F_\eta^\theta &=  \tfrac{1}{8}\Big(\gamma_2 (\xi-1)
+\tfrac{R}{2\chi^2}(\xi-2)\Big).
\end{align*}
The remaining lower order terms contained in~$S_{G^\pm},S_{B^\pm},$
and~$S_{U^\pm}$ are given in detail in appendix~\ref{Appendix}.  In
view of the definition of the reduction constraints in
equation~\eqref{ReductionConstraintToyModel} and the definition of the
rescaled variables~\eqref{eq:rescaling}, we define the rescaled
reduction constraints as
\begin{align*}
 {}^{(G)}\mathbb{C}_R  \equiv & \chi^2 ( {}^{(g)}\mathcal{C}_R),  &
 {}^{(G)}\mathbb{C}_A  \equiv & \chi ({}^{(g)}\mathcal{C}_A), \\
 {}^{(B)}\mathbb{C}_R  \equiv & \chi^2 ({}^{(b)}\mathcal{C}_R), &
 {}^{(B)}\mathbb{C}_A  \equiv & \chi ({}^{(b)}\mathcal{C}_A)
 + \tfrac{1}{8}(\xi-1)  ({}^{(\eta)}\mathcal{C}_A) , \\
 {}^{(U)}\mathbb{C}_R  \equiv & \chi^3 ({}^{(u)}\mathcal{C}_R),    &
 {}^{(U)}\mathbb{C}_A  \equiv & \chi^2 ({}^{(u)}\mathcal{C}_A).
\end{align*}
Notice that~${}^{(\eta)}\mathcal{C}_A$ is not rescaled as it is
associated with the auxiliary variable whose evolution equation was
chosen ad hoc. It is possible to define~$\eta$ so that it could be
treated on an equal footing with the other variables, but as it
already serves the purpose required we have not tried to do
so. Moreover, as the evolution equations~\eqref{eqn:evofinal} contain
at most first radial derivatives of~$\eta$ there is no need to
introduce a reduction variable to encode~$\p_r\eta$ and, consequently,
the associated reduction constraint~${}^{(\eta)}\mathcal{C}_R$ is also
not required. Direct computation using
equation~\eqref{ReductionConstraintToyModel} reveals,
\begin{align}
{}^{(G)}\mathbb{C}_{R}{} &= \tfrac{\chi}{R'}\p_rG
+ \tfrac{1}{2 R' c_{-}^{r}}G^{+}{} + \tfrac{ \chi}{2 R' c_{+}^{r}}G^{-}{}
- \tfrac{Q}{2 R' \chi c_{-}^{r}} G,\nonumber\\
{}^{(B)}\mathbb{C}_{R}{} &= \tfrac{\chi }{R'}\p_rB
-\tfrac{\chi (\xi-1) }{8 R'}\p_r\eta
+\tfrac{1}{2 R' c_{-}^{r}}B^{+}{} \nonumber\\
&\quad + \tfrac{ \chi}{2 R' c_{+}^{r}}B^{-}{} - \tfrac{Q}{2 R' \chi c_{-}^{r}}B
- \tfrac{R^2(\xi-1)}{8 R' \chi^5 c_{+}^{r}} P^2 \nonumber\\
&\quad+\tfrac{1}{16} \Bigl(2 \chi ' ( \xi -1)  - 2 \chi \xi'
+ \tfrac{R ( \xi-2 )}{R' \chi c_{-}^{r}}\Bigr)\eta, \nonumber\\
{}^{(U)}\mathbb{C}_{R}{} = \; &  \tfrac{\chi }{R'}\p_rU
+\tfrac{1}{2 R' c_{-}^{r}}U^{+}{} + \tfrac{ \chi}{2 R' c_{+}^{r}}U^{-}{}
-  \tfrac{Q}{R' \chi c_{-}^{r}} U,\label{eq:recaled_cons}
\end{align}
and
\begin{align*}
{}^{(G)}\mathbb{C}_A& = \p_A G-G_A, & {}^{(B)}\mathbb{C}_A &= \p_A B -B_A,\\
{}^{(U)}\mathbb{C}_A& = \p_A U-U_A, & {}^{(\eta)}\mathcal{C}_A &= \p_A \eta-\eta_A,
\end{align*}
where we have introduced~$Q=R+2 R'\chi\chi'c_{-}^{r}$ to write these
expressions succinctly, for the constraints. In the next subsection we
fix the asymptotics for~$H', R'$ and~$\chi$. Under those conditions
one can verify that~$Q\simeq Rc_{-}^{r}$.

\paragraph*{Discussion:} In~\cite{Luk13a}, in a mathematical relativity
context, the Gauss curvature of certain two-spheres was taken as an
unknown variable in place of a component of the four-dimensional
Riemann tensor, the two being related by the Gauss equation. In
hindsight our regularization strategy is rather similar, in the sense
that a nonlinear change of variables is made to try and derive
equations avoiding the worst behaved quantities. For now it is not
clear if this method can be applied to arbitrary PDEs satisfying some
kind-of weak-null condition, but we do suspect that to be the case.

\subsection{The compactified asymptotic system}\label{Section:CompactifiedAsymptoticSystem}

In this subsection we obtain the asymptotic expressions implied by the
evolution
equations~\eqref{eq:Recaled_Good_EoMs}-\eqref{eqn:evofinal}. Observe
that in these expressions neither the compression nor the height
functions~$R(r)$ and~$H(R)$ have been fixed, and we must now do so. In
the following we therefore consider~$\chi(R)$, $R(r)$ and~$H(R)$ with
the following asymptotics,
\begin{align}\label{asymptHprimeRprimeChiXi}
\chi \simeq R, \quad
\quad R'\simeq R^n, \quad H' \simeq 1 - R^{-n}.
\end{align}
Here~$1<n\leq2$ is a parameter that controls the asymptotic behavior
of~$R(r)$. Observe that the condition~$n>1$ is needed so that
$R\rightarrow \infty$ as $r\rightarrow r_{\mathscr{I}}$ for a
finite~$r_{\mathscr{I}}$. On the other hand, as discussed
in~\cite{CalGunHil05},~$0<n<2$ is required for numerical stability.
Near null-infinity the equations of motion then take the form,
\begin{align}\label{AsymptoticSystemGeneralFinal}
  \p_t\Psi^{+}{} &= -\tfrac{1}{2}\gamma_2\Psi^{+} +\mathcal{O}\Bigl(R^{n-2}
  (\ln R)^{p_{\Psi}}\Bigr) , \nonumber \\
  \p_t\Psi^{-}{} &=  -\p_r \Psi^{-} + \gamma_2\Bigl(-\p_r\Psi-\tfrac{1}{2}\Psi^-
  +\mathcal{O}\Bigl(R^{1-n}(\ln R)^{p_\Psi}\Bigr)\Bigr) \nonumber \\
&\quad+\mathcal{O}\Bigl(R^{n-2}(\ln R)^{p_{\Psi}}\Bigr) , \nonumber \\
  \p_t\Psi^{}{} &=\tfrac{1}{2}\Psi^{-} +\mathcal{O}\Bigl(R^{-1}
  (\ln R)^{p_{\Psi}}\Bigr) , \nonumber\\
\p_t\Psi_{A} &=\tfrac{1}{2}\p_A \Psi^- + \gamma_2\Bigl(\p_A \Psi-\Psi_A
-\tfrac{1}{8}p_\Psi(\xi-1)(\p_A \eta-\eta_A)\Bigr) \nonumber \\
&\quad+\mathcal{O}\Bigl(R^{-1}\Bigr) ,  \nonumber\\
\p_t \eta &=  (G^-){}^2 + \mathcal{O}(R^{-1}), \quad
\p_t \eta_A =  2 G^- \p_AG^- + \mathcal{O}(R^{-1}),
\end{align}
where~$\Psi \in \{ G,B,U \}$ with~$ p_G = p_U = 0 $ and~$ p_B = 1$.
These expressions suggest that, in order to obtain regular expressions
at~$\mathscr{I}^+$ it is necessary to assume some decay on~$\gamma_2$
and restrict the range of the parameter~$n$. Taking these
considerations into account, setting~$\gamma_2 \simeq R^{-\omega}$
with $\omega>0$ and taking~$1<n<2$, the asymptotic system reads
\begin{align}
  \p_{t}\Psi^{+}{} & \simeq  0,  &  \ell \;\Psi^{-}{}  & \simeq  0,
  & \p_{t}\Psi^{}{}  & \simeq  \; \tfrac{1}{2}\Psi^{-}, \nonumber\\
  \p_{t}\Psi_{A} & \simeq \; \tfrac{1}{2}\p_A \Psi^-,  &
  \p_t \eta \simeq & (G^-){}^2, &  \p_t \eta_A & \simeq 2 G^- \p_AG^-,
  \label{CleanAsymptoticSystem}
\end{align}
where~$\ell = \p_t + \p_r$. The latter vector corresponds,
asymptotically, to the outgoing null direction. To see this, a direct
computation using equation~\eqref{asymptHprimeRprimeChiXi} shows that
\begin{align}
\ell \simeq R^n L.
\end{align}
Similarly, the constraints take the form,
\begin{align*}
{}^{(\Psi)}\mathbb{C}_R & \simeq -\Psi^+ + R^{1-n}(\p_r \Psi
+ \tfrac{1}{2}\Psi^-) + \mathcal{O}(R^{(p_\Psi-n)}(\ln R)^{p_\Psi}),  \\
{}^{(\Psi)}\mathbb{C}_A & = \p_A \Psi-\Psi_A.
\end{align*}
The first of these implies that if the reduction constraints are
satisfied, then even if~$\p_r \Psi + \tfrac{1}{2}\Psi^-\simeq
\mathcal{O}(1)$, a condition weaker than that indicated by the
asymptotic system, then~$\Psi^{+} \simeq R^{1-n}$, and so it must
decay near~$\mathscr{I}^+$. It follows from the asymptotic
system~\eqref{AsymptoticSystemGeneralFinal} that the choice~$n=2$ is
out of reach for the bads if we insist on having regular equations for
regular unknowns. In the approach discussed in~\cite{VanHusHil14,
  VanHus17} the conformal factor is a fixed function of the radial
coordinate that regularizes the conformal metric and thus, in our
setup, corresponds to the choice~$n=2$. Consequently we can only
compare our good field with the wave equation in the setup
of~\cite{Van15}.  Although the asymptotic analysis has not been
performed for the formulation of~\cite{VanHusHil14} nor for the
conformal field equations~\cite{Fri81}, given what we have seen for
harmonic gauge the presence of logs in the solutions in those setups
is possible too. This point could be addressed by such an analysis for
those formulations, which we postpone for future work.

\section{Numerical Evolutions}

Having given the model and the strategy for regularization we now move
on to our numerical implementation. In section~\ref{Section:Code}
we discuss the methods employed and the specific data evolved. In
section~\ref{Section:Numerics_GBU} we present our data.

\subsection{Code overview}
\label{Section:Code}

\paragraph*{Continuum choices:} For the numerical implementation we take
the following for the height and compress functions,
\begin{align}
  H' = 1 - \frac{1}{R'}, \qquad R(r) = \frac{r}{\Omega^{\frac{1}{n-1}}},
  \qquad \Omega(r) = 1-\frac{r^2}{l^2},
\end{align}
so that~$\mathscr{I}^+$ is located at~$r=l$, and always
set~$l=1$. Observe that the above choice for the height function
implies that,
\begin{equation}
c_{+}^r = 1 \qquad c_{-}^r = -\tfrac{1}{2}R^{-n} +\mathcal{O}(R^{-2n}).
\end{equation}
The latter ensures that outgoing pulses propagate without distortion
as they move towards~$\mathscr{I}^+$. For the
rescaling function~$\chi$ and the damping parameter we choose
\begin{equation}\label{eq:chigamma2}
 \chi = \sqrt{1+R^2},  \qquad \gamma_2 = \tfrac{\gamma}{\chi} .
\end{equation}
This choice satisfies the conditions of
equation~\eqref{asymptHprimeRprimeChiXi} so that in the asymptotic
limit one recovers equations~\eqref{CleanAsymptoticSystem}.  The
reason for setting~$\chi = \sqrt{1+R^2}$ is to avoid introducing
unnecessary singular terms at the origin~$R=0$. As initial data
we set each of the raw fields~$g,b$ and~$u$, to
\begin{align}
a e^{-\delta (R-R_0)^2}+a e^{-\delta (R+R_0)^2},\label{eq:ID}
\end{align}
keeping the freedom to adjust the amplitude, width and offset
parameters~$a,\delta,R_0$ separately for each field. The values for
the regularized fields~$G,B$ and~$U$ are then computed by taking
derivatives and/or applying the change of variable in the obvious
manner. The auxiliary variable~$\eta$ is taken to vanish initially.

\paragraph*{Numerical setup:} Our experiments have been performed in a
one-dimensional code that uses very standard methods, and shares the
same basic infrastructure as that used for the spherically symmetric
hyperboloidal evolutions in~\cite{VanHusHil14,VanHus17}. We now give a
quick overview of these methods. The method of lines is employed for
time integration, and is performed with a fourth order accurate
Runge-Kutta. To approximate spatial derivatives we use second order
centered finite differences. We made this choice because the small
stencil makes the propagation of noise potentially slower than with
higher order finite differences or spectral methods. The only subtlety
in the implementation is that, because the evolution
equations~\eqref{eq:firstorderevolutionequationslowercasecoords} were
written in spherical polar coordinates the
equations~\eqref{eq:Recaled_Good_EoMs}-\eqref{eqn:evofinal} contain
divergent terms at the origin. Since we are performing spherically
symmetric evolutions the~$1/R^2$ coefficient
of~$q^{AB}\slashed{\nabla}_A\Psi_B$ is not problematic, but the~$1/R$
terms present in the source terms~$S_\Psi$ (displayed in
Appendix~\ref{Appendix}) require special attention. Our solution is to
use Evans method~\cite{Eva84a} as discussed in~\cite{GunGarGar10} in
the context of summation by parts discretizations of the wave equation
in spherical symmetry. In the latter it is shown that given a system
of equations of the form,
\begin{align}
\p_t\psi = \p_r \pi, \qquad \p_t \pi = \p_r\psi + \tfrac{p}{r}\psi ,
\end{align}
the spatial derivatives can be discretized as,
\begin{align}
\p_t\psi = h^{-1} D \pi, \qquad \p_t \pi = h^{-1}\tilde{D}\psi,\label{eq:Evans}
\end{align}
where~$h \equiv \Delta r$ is the grid spacing and the difference
operators~$D$ and~$\tilde{D}$ are given by
\begin{align*}
h^{-1}\tilde{D}\psi & = (p+1) \frac{r^p_{i+1}\psi_{i+1}-r^p_{i-1}
\psi_{i-1}}{r^{p+1}_{i+1}-r^{p+1}_{i-1}},\\
h^{-1}D\pi &= \frac{\pi_{i+1}-\pi_{i-1}}{2},
\end{align*}
where~$\psi_i(t)$ and~$\pi_i(t)$ are grid functions
approximating~$\psi(t,r)$ and~$\pi(t,r)$ on a grid~$r_i$. We work
always with a non-staggered grid so that there are gridpoints both
directly at the origin and at~$\mathscr{I}^+$. In order to rewrite
equations~\eqref{eq:Recaled_Good_EoMs}-\eqref{eqn:evofinal} in a form
in which the discretization~\eqref{eq:Evans} can be applied, we define~$\Xi$
according to,
\begin{equation}
\frac{1}{R} = \frac{1}{R'}\Big( \frac{1}{r} - \frac{\Xi'}{\Xi}\Big),
\end{equation}
which can be used to rewrite the terms in~$S_\Psi$ with contain~$1/R$
and then exploit the aforementioned discretization to absorb the
singular behavior at the origin with~$D$ and~$\tilde{D}$. Although
these equations are regular everywhere (including at~$\mathscr{I}^+$),
the coefficients in these equations are in general of the
form~$Q_m(R)/Q_{s}(R)$ with~$m \leq s$ or~$\ln(R)Q_m(R)/Q_{s}(R)$
with~$m < s$ where~~$Q_n(R)$ denotes a polynomial in~$R$
of degree~$n$. Thus to avoid evaluating numerically the quotient of
two large numbers careful algebraic manipulations are required. In
practice, one can opt also to substitute~$R(r)$ explicitly. To manage
the inner boundary we define ghostzones, which are populated from the
bulk variables using the known parity of the raw~$g,b$ and~$u$ fields
and their derivatives. By construction no physical boundary conditions
are needed at~$\mathscr{I}^+$, but derivatives must still be
approximated. To make that possible we extrapolate the evolved fields
from the bulk into one ghostzone at fourth order and use the standard
spatial operators all of the way out to the boundary point. The final
ingredient in our method is the use of Kreiss-Oliger
dissipation~\cite{KreOli73}
\begin{align}
\sigma h^3 D^2_+D^2_-/16\,,\label{eq:KO}
\end{align}
with~$D_\pm$ the standard forward and backward differencing operators,
which, as used in~\cite{BabHusHin07}, is added to each of the
evolution equations to reduce high-frequency noise. An exception is
the~$\eta$ variable, which is treated differently because no spatial
derivatives of this quantity are present anywhere in the system, and
experimentally we find that this leads to a misleading third order
convergent feature at the resolutions we employ when dissipation is
used on the variable. At the outer boundary we use the same
extrapolation mentioned above to fill the additional point in the
dissipation stencil.

\subsection{Results with the GBU model}
\label{Section:Numerics_GBU}

\begin{figure*}[t]
\centering
\includegraphics[width=0.45\textwidth]{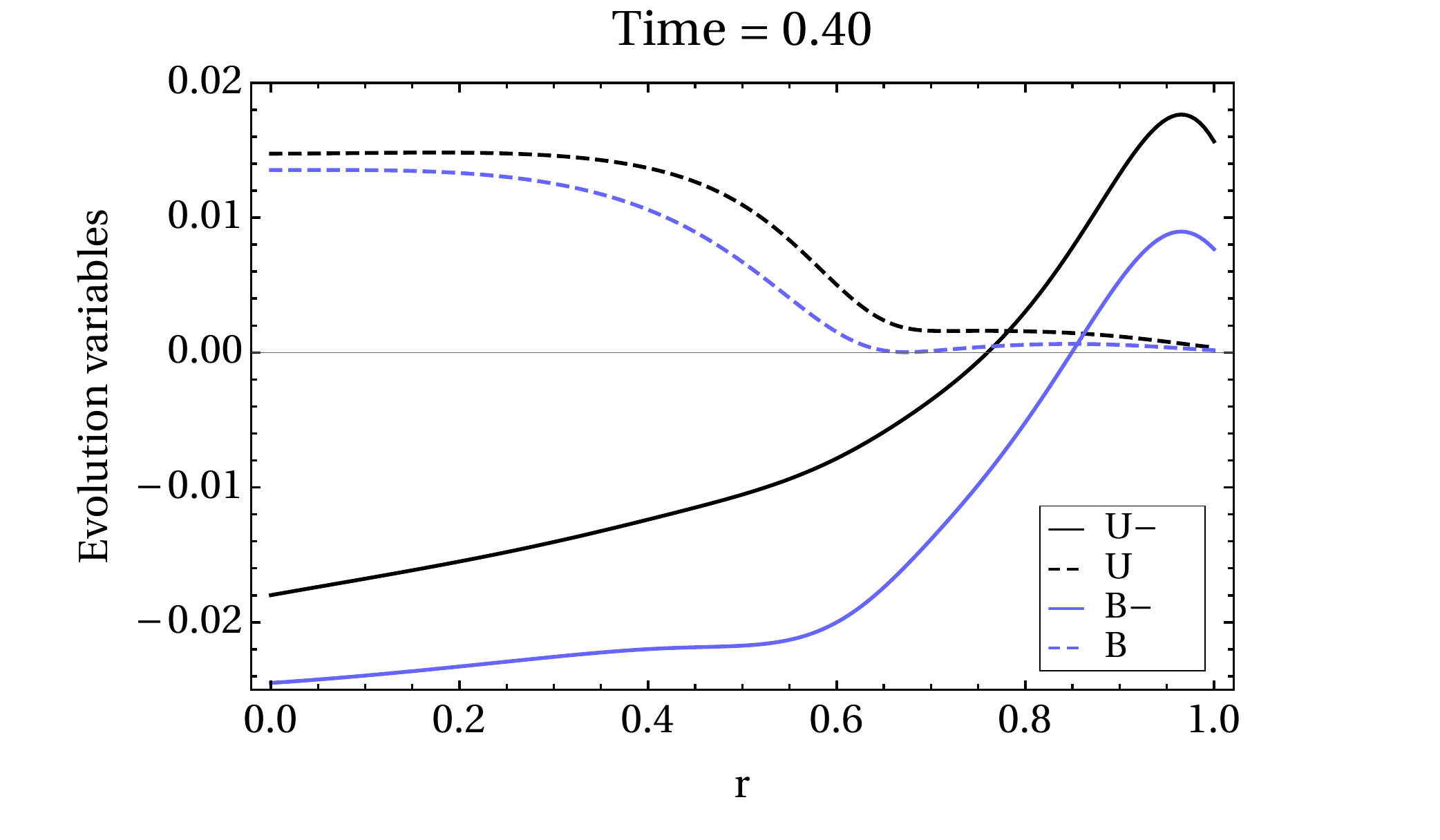}
\includegraphics[width=0.45\textwidth]{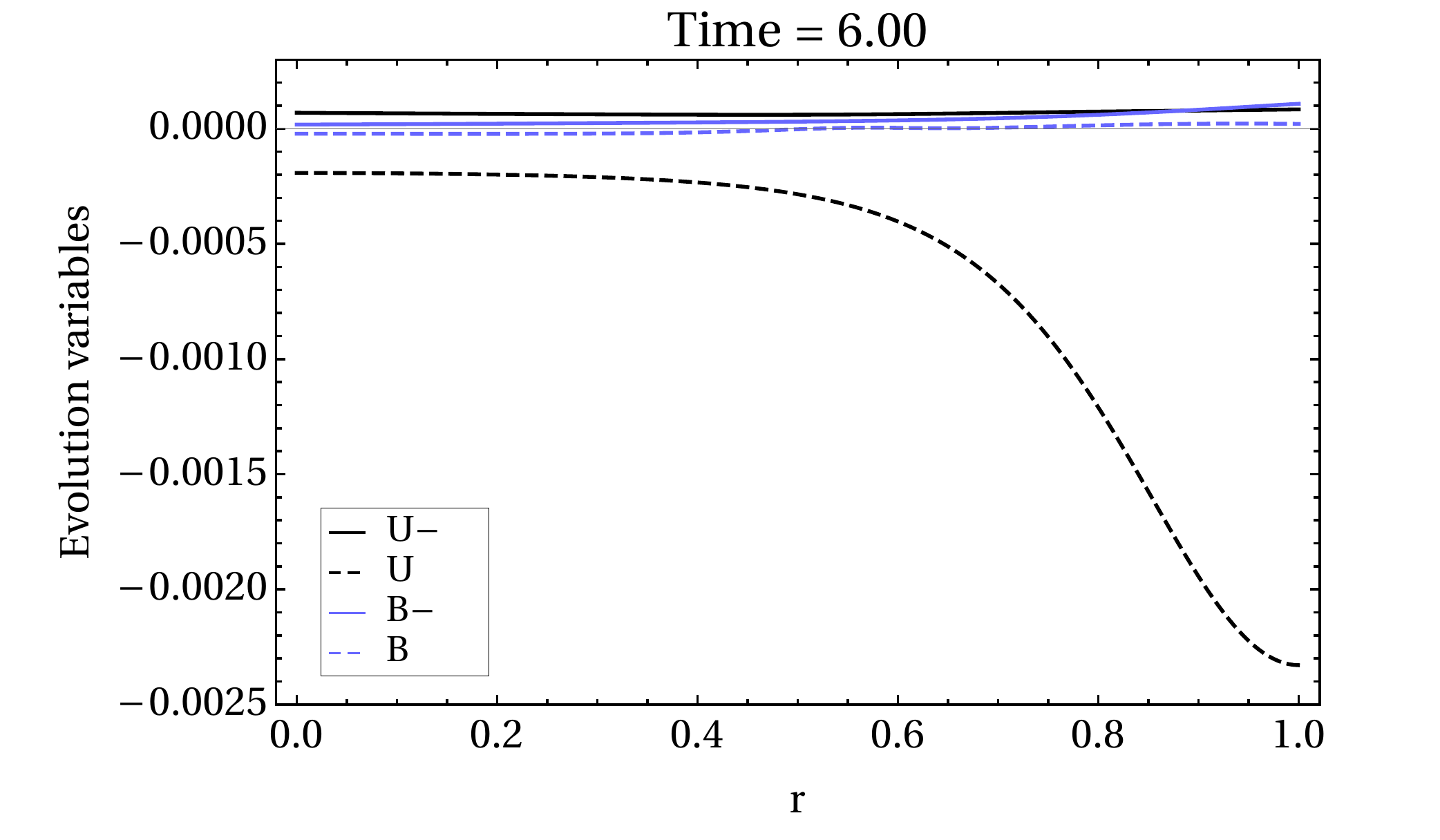}
\caption{In these plots we display snapshots of the solution for the
  regularized ugly~$U^-, U$ and bad~$B^-, B$ fields at fixed times,
  against the compactified radial coordinate~$r$. In the left panel
  the time~$t=0.4$ was chosen as it corresponds roughly with the
  maximum of the radiation field~$B^-$ at null-infinity (compare with
  Fig.~\ref{fig:B-2D}).
  The right panel serves to demonstrate the
  decay of the fields after the initial pulse leaves the domain. Note
  the difference in scale.}
  \label{fig:Snapshots}
\end{figure*}

\begin{figure*}[t]
  \includegraphics[width=0.45\textwidth]{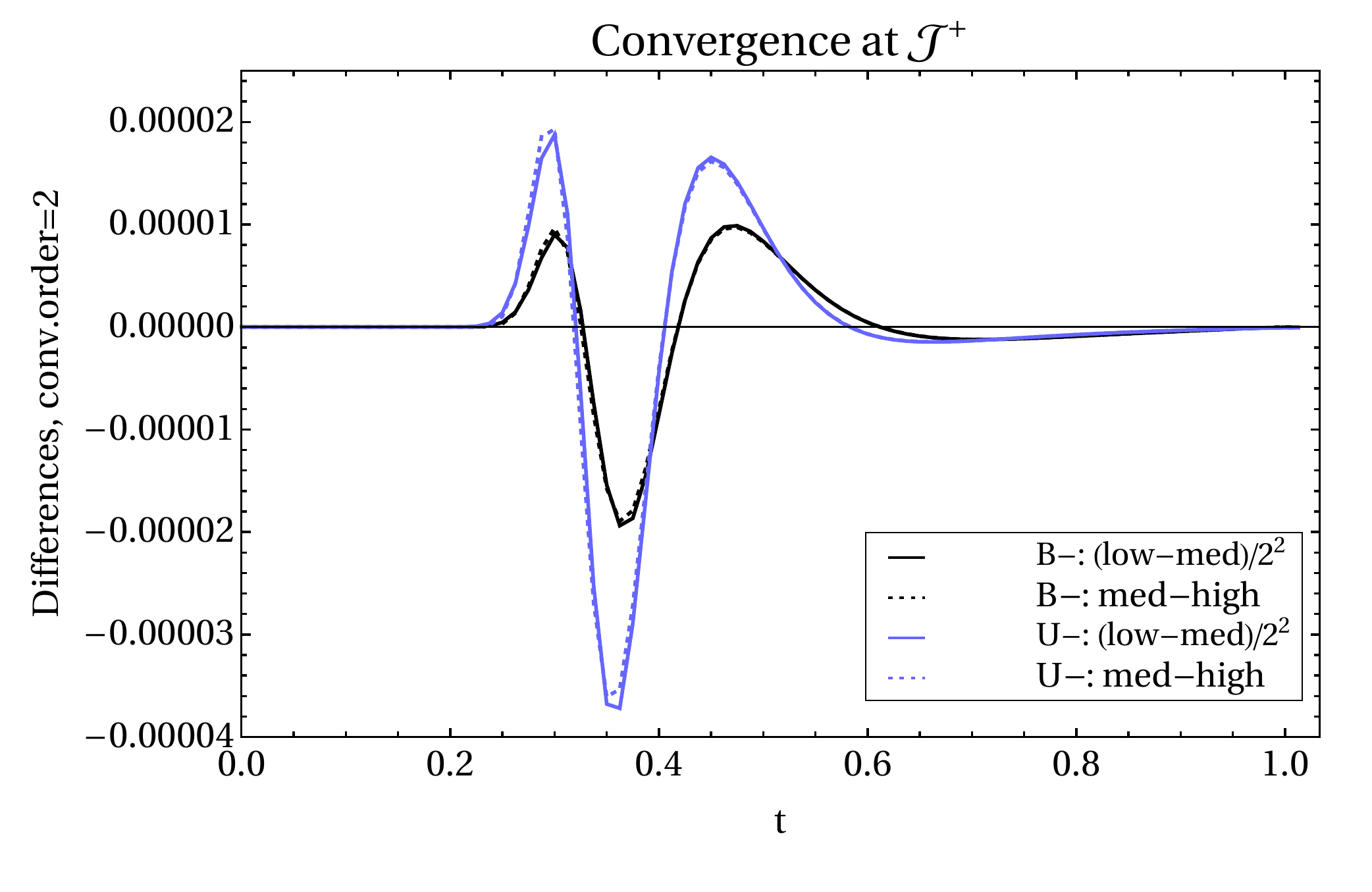}
  \includegraphics[width=0.475\textwidth]{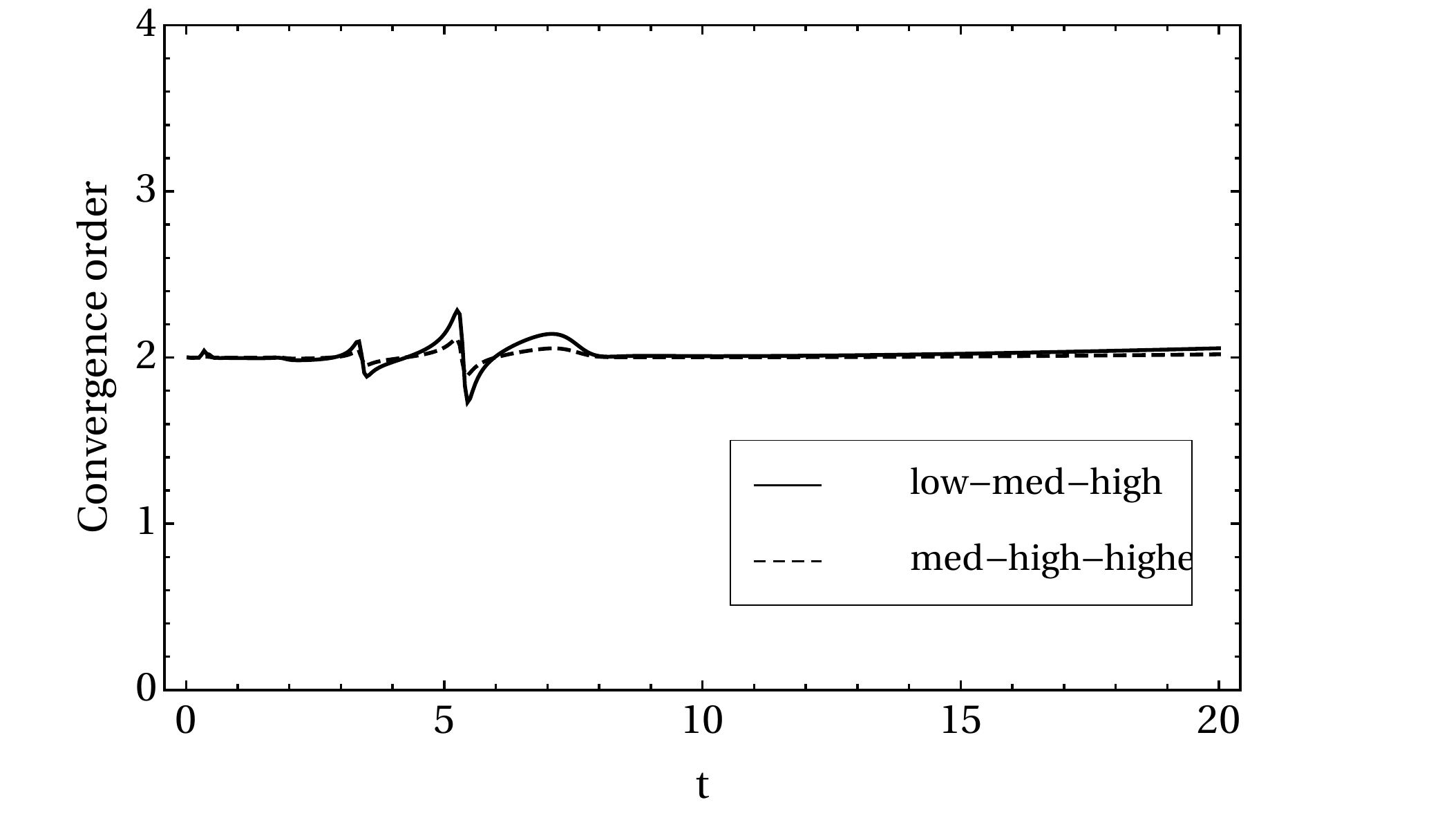}
  \caption{To judge whether or not the numerical method is working
    reliably we perform convergence tests as usual. In going from low
    to medium and medium to high resolution we doubled resolution. In
    the left panel we demonstrate that the outgoing radiation
    fields~$U^-$ and~$B^-$ converge perfectly at second order at
    null-infinity as a function of time, consistent with our
    discretization. In the second panel we display the experimental
    convergence rate in the norm of the difference of the solutions as
    a function of time. At first glance it seems that there is a
    problem abound~$t=4$ and~$t=6$, but, as can be seen by the second
    curve, this effect converges away.}
  \label{fig:Conv_Plots}
\end{figure*}

\paragraph*{Basic dynamics:} We begin with a description of the basic
dynamics of the system. To give ball-park figures we find that~$200$
spatial points and a CFL factor of~$1/2$ are typically sufficient to
provide well-resolved solutions that appear smooth in space and time
given our initial data, and is also sufficient to see convergence
experimentally, see Fig.~\ref{fig:Conv_Plots}. We work always with the
dissipation parameter~$\sigma=2/100$,
see~\eqref{eq:KO}. Following~\eqref{eq:ID} and placing identical data
centered at the origin~$R_0=0$ with width parameter~$\delta=1$ and
amplitude~$a=1/100$ in each of the raw fields~$g,b,u$, and trivial
initial data for the auxiliary variable~$\eta$, we see the expected
behavior. The data for each field splits into two pulses. The first of
these propagates directly out to null-infinity, whereas the second
appears to oscillate briefly at the origin first. A brief comparison
with results of an older code for the plain-wave equation in
first order in time, second order in space form, written in the same
infrastructure, against the good field~$g$ reveals comparable results,
again giving us confidence in the numerics (taking~$n=2$ as mentioned
earlier). In figure~\ref{fig:Snapshots} we plot snapshots of the
solution for the bad and ugly fields given the aforementioned initial
data. The ugly field~$u$ is decoupled from the rest of the system, but
behaves in a qualitatively similar manner. At late times we see a hump
near~$\mathscr{I}^+$ in the rescaled ugly field~$U$ that appears to
decay slowly. The important first result here is that the behavior
predicted by the asymptotic system and hoped for in our regularization
is realized; our evolved fields and their derivatives are finite, and
the equations of motion are explicitly regular. The basic dynamics
described above does not change if one adds a small offset~$R_0\neq0$
or gives different widths~$\delta$ for each of the raw fields $g,b,u$.

\begin{figure}[t]
  \centering
  \includegraphics[width=0.45\textwidth]{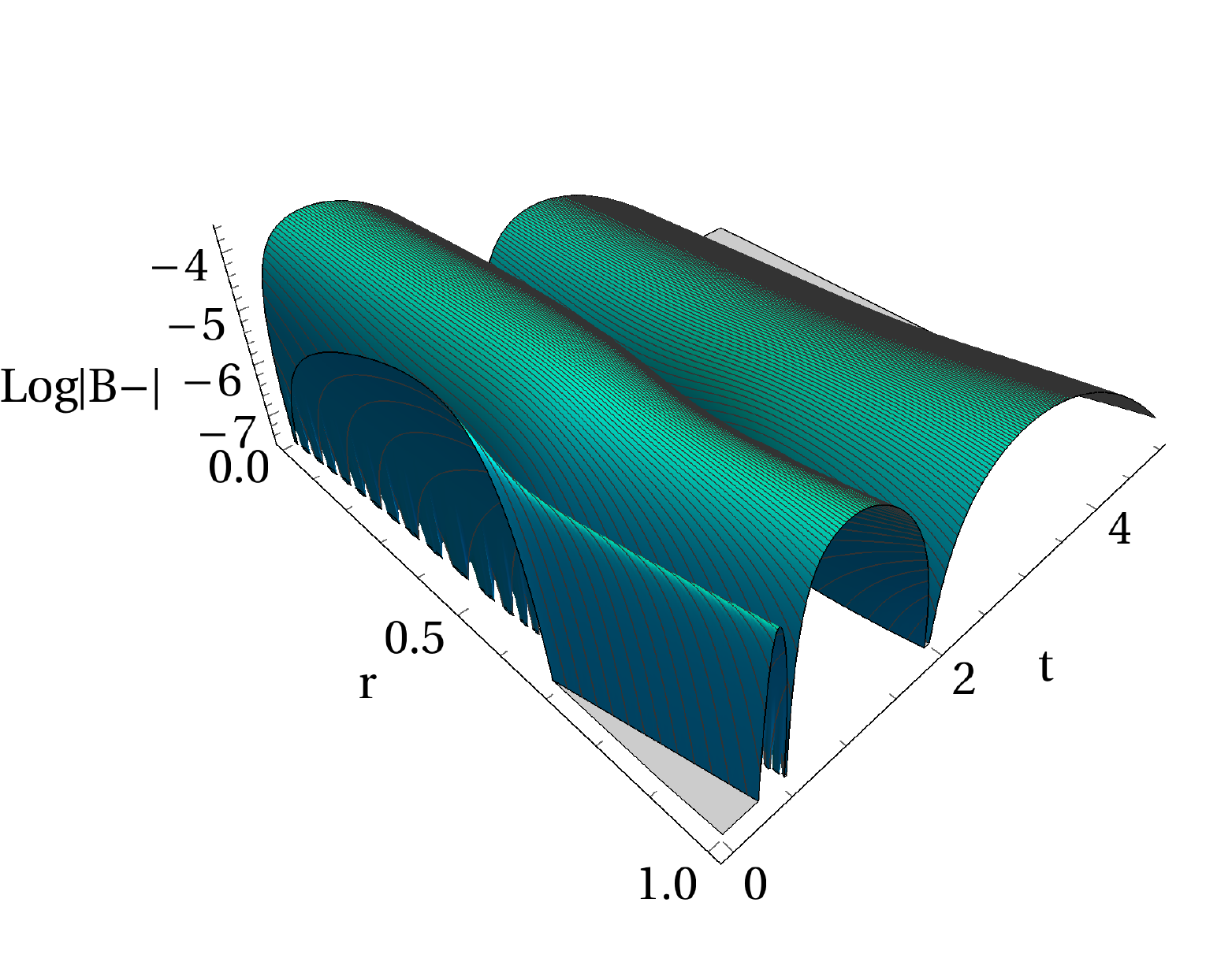}
  \caption{Here we plot the reformed bad field~$B^-$ as a function of
    spacetime, again against the compactified radial
    coordinate~$r$. The take home message is that the initial pulse
    exits the domain cleanly, with no visible numerical reflection,
    leaving behind a low-frequency feature in space that gradually
    decays.}
  \label{fig:B-2D}
\end{figure}

\paragraph*{Behavior of the reformed bad and auxiliary fields:} For our
particular model there is no question of finite time blow-up, so any
explosion of the data must be caused by a failure of the numerical
method; we see no such blow-up. That said, as we increase the
amplitude of the~$g$ field or give non-trivial initial data
for~$\eta$, we see both that the distortion in~$B$ increases and,
obviously, that~$\eta$ grows, indicating that the original~$b$
variable is picking up a log-term as expected. In
figure~\ref{fig:B-2D} we display a spacetime plot of the outgoing
radiation-field~$B^-$. Since the~$g$ field rapidly propagates out
through~$\mathscr{I}^+$, we find that~$\eta$ reaches a fixed, non-zero
end-state rather quickly.

\paragraph*{Constraint damping:} One potential weakness of the present
formulation is that for regularity of the field equations we have to
suppress the constraint damping parameter~$\gamma_2$ like~$1/R$
near~$\mathscr{I}^+$. One might therefore worry that enforcing
strong-damping~$\gamma_2=O(1)$ near the origin would result in small
violations near the origin and large violations near~$\mathscr{I}^+$,
which could generate large gradients and hence large errors, even if
the scheme were converging reliably.  To
investigate this we have compared evolutions with and without
constraint damping switched on, and find that within the run-times
considered~$t\simeq 100$, such problems do not manifest. We think it
may be possible to adjust the present constraint damping scheme to
damp violations on outgoing pulses in such a way that we could
maintain~$O(1)$ constraint damping parameters, but since doing so
would require re-engineering the entire scheme, and we presently do not
see a pressing need for such a modification, we have not pushed this
line of inquiry further.

\paragraph*{Convergence:} In the future we hope that a regularization
similar to that employed here will be useful in gravitational wave
applications. Such work requires meaningful error-estimates. Our
experiments can therefore only be considered a success if clean
convergence can be experimentally achieved. To investigate this we
performed a set of runs in which we start from the grid-setup
mentioned above and then doubled resolution several times, keeping the
remaining parameters fixed. These tests were furthermore performed
with several different choices for the initial data parameters. In
figure~\ref{fig:Conv_Plots} we display two of the
ensuing convergence plots. In the first we show that perfect pointwise
convergence for the outgoing radiation fields~$B^-$ and~$U^-$ is
attained at~$\mathscr{I}^+$. The second shows that good second order
convergence is obtained in the norm of the full solution as resolution
is increased. In the latter we do notice a gradual drift away from
second order at late times, but this effect is suppressed as
resolution increases, so does not appear to be a problem {\it in
  principle}. It may be that by adjusting the specifics of our outer
boundary treatment this behavior can be improved. So far we have not
done so systematically however. We have also examined the constraint
violations~\eqref{eq:recaled_cons} and find perfect second order
convergence in all quantities.

\section{Conclusions}

Continuing towards a robust treatment of future null-infinity in
numerical relativity we considered a semilinear system of wave
equations. The system was constructed with the nonlinear structure of
the field equations of GR in harmonic gauge in mind, and has three
different types of fields. We call these the good, the bad and the
ugly. Of these, the good fields have fall-off near null-infinity like
solutions of the wave equation, whereas the ugly fields decay
faster. Finally the bad fields decay worse that solutions to the wave
equation by a logarithm in~$R$. The main accomplishment in this paper
was to give a reformulation of the equations that delivers regular
equations for regular, generically non-decaying, variables on
compactified hyperboloidal slices. The regularization strategy is to
rescale all of the wave fields as aggressively as possible and then
define new fields to subtract out any potential logarithmic, or
perhaps harsher, divergences. The crucial technical tool was the use
of an asymptotic expansion, which allowed us to discard irrelevant
terms. In our model this meant the introduction of the~$\eta$ field,
and by analogy in GR will mean integrating up the square of the Bondi
news function near null-infinity. Remarkably the asymptotic expansion
of the regularized system is identical to that of a set of decoupled
wave equations.  With the regularization in hand we presented a set of
numerical evolutions in which we demonstrated perfect pointwise and
norm convergence over the entire computational domain, which included
null-infinity explicitly. Future tests of the model will be performed
using the~\texttt{bamps} pseudospectral numerical relativity
code~\cite{Bru11,HilWeyBru15}, which presents a more subtle challenge
because spectral methods may more susceptible to any lingering
lower-order log-terms. In forthcoming work we will present a
regularization of GR following the same approach.

\acknowledgments

We are grateful to Sukanta Bose, Bernd Br\"ugmann, Rodrigo Panosso
Macedo, Isabel Su\'arez Fern\'andez and Juan Valiente Kroon for
helpful discussions or comments on the manuscript. EG, DH and AVV also
gratefully acknowledge the support and hospitality offered by IUCAA,
Pune, where part of this work was completed. Many of our derivations
were performed in xAct~\cite{xAct_web_aastex} for Mathematica. The
notebooks are available at~\cite{GasHil19_web}. The work was partially
supported by the FCT (Portugal) IF Program~IF/00577/2015,
PTDC/MAT-APL/30043/2017, the European Research Council Consolidator
Grant 647839, the GWverse COST action Grant No. CA16104, and under the
PhD researcher Decree-Law no.~$57/2016$ of August~$29$ (Portugal),
IUCAA's NRTT grant and the UGC of India.

\appendix

\section{Source terms in the GBU model}\label{Appendix}

The source terms~$S_\Psi$ of equation~\eqref{eqn:evofinal} are expressions
polynomials in the fields~$\Phi$ whose coefficients are regular
functions of~$R$. Using the notation of
Section~\ref{Section:EvoEqsForTheRescaledVariables}, the sources
for~$G^+$ and~$G^-$ are given by
\begin{align*}
&S_{G^+}= - \tfrac{R' c_{-}^{r}}{R\chi}(R -  \chi) (R + \chi) G^{-}
 +\tfrac{2R' c_{-}^{r}}{\chi^2}(\chi - R \chi ') G\\
&+ \tfrac{\gamma_2}{2}\Bigl(\tfrac{1}{\chi} (R + 2 R' \chi \chi ' c_{-}^{r}) G
 - G^{+}{} - \chi (1 + 2 R' c_{-}^{r})G^{-}{}\Bigr)\\
 &+\tfrac{R'c_{-}^{r}}{R \chi}(2 R \chi'- \chi)G^{+},
\end{align*}
and
\begin{align*}
S_{G^-} &= \tfrac{R'c_{+}^{r}}{R\chi}G^{+}
 + \tfrac{R' c_{+}^{r}}{R\chi} (R \chi'- \chi)G^{-}-\tfrac{ R'c_{+}^r}{\chi^2}G\\
&+ \tfrac{\gamma_2}{2}\Bigl(\tfrac{1}{\chi}(2 R' c_{+}^{r}-1) G^{+}-G^{-}{} \\
&+\tfrac{1}{\chi^2}(R - 2 R' R c_{+}^{r}
 + 2 R' \chi \chi ' c_{+}^{r})G\Bigr).
\end{align*}
For the $B^+$ and $B^-$ fields we have,
\begin{align*}
&S_{B^+} = \tfrac{1}{8\chi^7}F_{1}^{+}(\chi^2G^{-}-\chi G^+ +RG)^2
 - \tfrac{R' c_{-}^{r}}{8\chi^2}F_{3}^{+} \eta\\
&+\tfrac{\gamma _2}{2}\Bigl(\tfrac{1}{8\chi} F_{2}^{+}\eta-B^+
 +\tfrac{1}{\chi}(R + 2 R' \chi  \chi ' c_{-}^{r})B\\
&+\tfrac{R^2c_{-}^r}{4\chi^5c_{+}^{r}}(\xi-1)(\chi^2G^{-}-\chi G^++RG)^2
 -\tfrac{\chi c_{-}^{r}}{c_{+}^{r}}B^-\Bigl)\\
  &-\tfrac{R'c_{-}^{r}}{R\chi}(R  - \chi ) (R + \chi ) B^-
 + \tfrac{2R' c_{-}^{r}}{ \chi^2}(\chi   - R \chi ')B\nonumber\\
&+\tfrac{R'c_{-}^{r}}{R\chi}(2 R \chi '-\chi) B^+,\\
 &S_{B^-} = \bigl(\tfrac{R'R}{2\chi^6}F_{1}^{-}
 - \tfrac{\gamma_2 R^2c_{-}^{r}}{4\chi^4c_{+}^{r}}( \xi-1)F_{7}^{-}\bigr)GG^{-}
 + \tfrac{R' c_{+}^{r}}{8\chi^2}(\xi-2)\eta\\
 &
 + \tfrac{\gamma_2c_{+}^r F_{9}^{-}}{16\chi^2c_{-}^{r}}\eta
 +  \bigl(\tfrac{R'c_{+}^r}{R\chi}(R \chi '- \chi)-\tfrac{\gamma_2}{2}\bigr)B^-
 + (\tfrac{R'c_{+}^{r}}{R\chi} - \tfrac{\gamma_{2}c_{+}^{r}}{2\chi c_{-}^{r}})B^{+}\\
 &+\bigl(\tfrac{\gamma_2R^2c_{-}^{r}}{8\chi^{2}c_{+}^{r}} (\xi-1)
 (2 R' c_{+}^{r} -3)-\tfrac{R'}{4 \chi^4}F_{3}^{-}\bigr) (G^-)^2\nonumber\\
 &+ \bigl(\tfrac{\gamma_2 R^2}{8\chi^4}(\xi-1) (4 R' c_{+}^{r}-3)
 -\tfrac{R'}{4\chi^6}(F_{4}^{-} + F_{5}^{-})\bigr)(G^{+}{})^2\\
 &- \bigl(\tfrac{R' c_{+}^r}{\chi^2}
  - \tfrac{\gamma_{2}}{2 \chi^2} (R - 2 R' R c_{+}^{r}
 + 2 R' \chi \chi ' c_{+}^{r})\bigr) B\\
 &+ \bigl(\tfrac{\gamma_{2}R^3c_{-}^{r}F_{8}^{-}}{8\chi^6c_{+}^{r}}(\xi-1)
  - \tfrac{R'R^2}{4 \chi^8}(F_{4}^{-} + F_{5}^{-}
  - 4 R' R \chi^2 c_{-}^{r} c_{+}^{r})\bigr) G^2\\
  &+\bigl(
  \tfrac{R'R}{2\chi^7}(F_{4}^{-} + F_{5}^{-} - 2 R' R \chi^2 c_{-}^{r} c_{+}^{r})
  -\tfrac{\gamma_2 R^2 c_{-}^{r}F_{6}^{-}}{4 \chi^5c_{+}^{r}} (\xi-1)\bigr) GG^{+}\\
  &-\bigl(\tfrac{ R'F_{2}^{-}}{2\chi^5}
  +\tfrac{\gamma_{2} R^2 c_{-}^{r}}{4\chi^3c_{+}^r}(\xi-1)
  (3 - 6 R' c_{+}^{r} + 4 R'^2 (c_{+}^{r})^2)\bigr)G^{-}G^{+},
\end{align*}
where the coefficients~$F^{\pm}$ are functions of~$R$ only. The
detailed expressions for these are given in the following
lists. The~`$+$' quantities are,
\begin{align*}
  F_{1}^{+} &= R^3 - 2 R' R^3 c_{-}^{r} + 2 R' R \chi^2 c_{-}^{r}
  + 2 R' \chi^3 c_{-}^{r}\\
  &\quad + 2 R' R^3 \xi c_{-}^{r}- 2 R' R \chi^2 \xi c_{-}^{r},\\
  F_{2}^{+} &= 2 R -  R \xi + 2 R' \chi \chi ' c_{-}^{r}
  - 2 R' \chi \chi ' \xi c_{-}^{r} + 2 R' \chi^2 \xi ' c_{-}^{r},\\
  F_{3}^{+} &= -4 \chi + 5 R \chi ' + 2 \chi \xi - 2 R \chi ' \xi.
\end{align*}
The~`$-$' quantities are,
\begin{align*}
F_{1}^{-} &= \chi^3 c_{+}^{r} - 2 R^3 (\xi-1) c_{-}^{r} (R' c_{+}^{r}-1)\\
&\quad+ R \chi^2 (\xi-1) c_{-}^{r} (8 R'c_{+}^{r}-3)\\
&\quad - R^2 \chi \bigl(\chi \xi ' c_{+}^{r}
+ \chi ' (\xi-1) c_{-}^{r} (6 R' c_{+}^{r}-1)\bigr),\\
F_{2}^{-} &= \chi^3 c_{+}^{r} - 3 R \chi^2 (\xi-1) c_{+}^{r}
- 2 R^3 ( \xi-1) c_{-}^{r} (R' c_{+}^{r}-1)\\
&\quad-R^2 \chi \bigl(\chi \xi ' c_{+}^{r}
+ \chi ' ( \xi-1) c_{-}^{r} (6 R' c_{+}^{r}-1)\bigr),\\
F_{3}^{-} &= 2 R^3 ( \xi-1) c_{-}^{r} + \chi^3 c_{+}^{r}
- R^2 \chi (\chi ' -  \chi ' \xi + \chi \xi ') c_{+}^{r}\\
&\quad+ R \chi^2 ( \xi-1) c_{-}^{r} ( 2 R' c_{+}^{r}-3),\\
F_{4}^{-} &= - R \chi \bigl(R \chi \xi '
- 10 R' \chi (\xi-1) c_{-}^{r}
+ 10 R' R \chi ' ( \xi-1) c_{-}^{r}\bigr) c_{+}^{r}, \\
F_{5}^{-} &= -3 R \chi^2 ( \xi-1) c_{-}^{r}
+ R^2 \chi \chi ' ( \xi-1) c_{-}^{r} + \chi^3 c_{+}^{r}\\
&\quad+ 2 R^3 (\xi-1) c_{+}^{r},\\
F_{6}^{-} &= 4 R' \chi \chi' c_{+}^{r} ( R' c_{+}^{r}-1)
+ R ( 10 R' c_{+}^{r} - 8 R'^2 (c_{+}^{r})^2 -3),\\
F_{7}^{-} &= 4 R' \chi \chi ' c_{+}^{r} ( R' c_{+}^{r}-1)
+ R ( 6 R' c_{+}^{r} - 4 R'^2 (c_{+}^{r})^2 -3),\\
F_{8}^{-} &= 8 R' \chi \chi ' c_{+}^{r} ( R' c_{+}^{r}-1)
+ R ( 10 R' c_{+}^{r} - 8 R'^2 (c_{+}^{r})^2 -3),\\
F_{9}^{-} &= - R ( \xi-2) + 2 R' \chi (\chi ' -  \chi ' \xi + \chi \xi ') c_{-}^{r}.
\end{align*}
Finally, the sources for~$U^\pm$ are given by,
\begin{align*}
S_{U^+} &= \tfrac{R'c_{-}^{r}}{R\chi}(\chi-R) (2 R + \chi)U^{-}
+ \tfrac{2R' c_{-}^{r}}{\chi^2} (2 \chi- R - 2 R \chi ')U\\
&- \tfrac{\gamma_{2}}{2} \Bigl(U^{+}{} + \chi (1 + 2 R' c_{-}^{r})U^{-}{}
- \tfrac{2}{\chi} (R + 2 R' \chi \chi ' c_{-}^{r}) U  \Bigr)\\
&+\tfrac{R'}{R\chi c_{-}^{r}} (R -  \chi + 3 R \chi ') U^{+},
\end{align*}
and,
\begin{align*}
S_{U^-} &= \tfrac{2 R' c_{+}^{r}}{ \chi^3}(R -  \chi)U
-\tfrac{R'}{R\chi^2c_{+}^{r}}(R -  \chi)U^{+}\\
&-\tfrac{R'c_{+}^{r}}{R\chi}(R + \chi - 2 R \chi') U^{-} \\
&-\tfrac{\gamma_2}{2}\Bigl(U^{-}{} -\tfrac{1}{\chi} (2 R' c_{+}^{r}-1) U^{+}\\
&- \tfrac{2}{\chi} (R - 2 R' R c_{+}^{r} + 2 R' \chi \chi ' c_{+}^{r})U \Bigr).
\end{align*}

\bibliographystyle{unsrt}
\bibliography{WeakNullToy.bbl}{}

\end{document}